\documentclass[lettersize,journal]{IEEEtran}
\usepackage{amsmath,amsfonts}

\usepackage{array}
\usepackage[caption=false,font=normalsize,labelfont=sf,textfont=sf]{subfig}
\usepackage{textcomp}
\usepackage{stfloats}
\usepackage{url}
\usepackage{verbatim}
\usepackage{graphicx}
\usepackage{cite}
\usepackage{hyperref}
\usepackage[right]{lineno}
\usepackage[table]{xcolor}

\usepackage{xcolor}

\usepackage[linesnumbered,ruled,vlined]{algorithm2e}
\usepackage{setspace}
\usepackage[noend]{algpseudocode}
\usepackage{environ}
\usepackage{textcomp}
\usepackage{soul}
\usepackage{filecontents}
\usepackage{tabularx}
\usepackage{adjustbox}
\usepackage{algpseudocode}
\usepackage{csquotes}

\usepackage{filecontents,catchfile}
\usepackage{subfiles}
\usepackage{xr-hyper}

\usepackage{xr}
\makeatletter

\newcommand*{\addFileDependency}[1]{
\typeout{(#1)}
\@addtofilelist{#1}
\IfFileExists{#1}{}{\typeout{No file #1.}}
}\makeatother

\newcommand*{\myexternaldocument}[1]{%
\externaldocument{#1}%
\addFileDependency{#1.tex}%
\addFileDependency{#1.aux}%
}

\myexternaldocument{Supp-info}

\begin{document}

\title{CLOG-CD: Curriculum Learning based on Oscillating Granularity of Class Decomposed Medical Image Classification}

\author{Asmaa Abbas, Mohamed Medhat Gaber, and Mohammed M. Abdelsamea
\thanks{A. Abbas and M. M. Gaber are with School of Computing and Digital Technology, Birmingham City University, Birmingham, UK, and M. M. Abdelsamea is with computer science department, University of Exeter, Exeter, UK.}
}

\markboth{Journal of \LaTeX\ Class Files,~Vol.~14, No.~8, August~2021}%
{Shell \MakeLowercase{\textit{et al.}}: A Sample Article Using IEEEtran.cls for IEEE Journals}


\maketitle

\begin{abstract}
Curriculum learning strategies have been proven to be effective in various applications and have gained significant interest in the field of machine learning. It has the ability to improve the final model's performance and accelerate the training process. However, in the medical imaging domain, data irregularities can make the recognition task more challenging and usually result in misclassification between the different classes in the dataset. Class-decomposition approaches have shown promising results in solving such a problem by learning the boundaries within the classes of the data set. In this paper, we present a novel convolutional neural network (CNN) training method based on the curriculum learning strategy and the class decomposition approach, which we call \emph{CLOG-CD}, to improve the performance of medical image classification. We evaluated our method on four different imbalanced medical image datasets, such as Chest X-ray (CXR), brain tumour, digital knee x-ray, and histopathology colorectal cancer (CRC). \emph{CLOG-CD} utilises the learnt weights from the decomposition granularity of the classes, and the training is accomplished from descending to ascending order (i.e. anti-curriculum technique). We also investigated the classification performance of our proposed method based on different acceleration factors and pace function curricula. We used two pre-trained networks, ResNet-50 and DenseNet-121, as the backbone for \emph{CLOG-CD}. The results with ResNet-50 show that \emph{CLOG-CD} has the ability to improve classification performance with an accuracy of 96.08\% for the CXR dataset, 96.91\% for the brain tumour dataset, 79.76\% for the digital knee x-ray, and 99.17\% for the CRC dataset, compared to other training strategies. In addition, with DenseNet-121, \emph{CLOG-CD} has achieved 94.86\%, 94.63\%, 76.19\%, and 99.45\% for CXR, brain tumour, digital knee x-ray, and CRC datasets, respectively.
\end{abstract}


\section{Introduction}
\label{sec:introduction}
\IEEEPARstart{M}{}edical imaging has made a significant contribution to the advancement of medicine and especially to the early detection of various diseases. For example, CXR is one of the most widely used medical imaging technology, and it is critical for diagnosing many thoracic diseases such as pneumonia, Covid-19, and lung nodule \cite{apostolopoulos2020covid}. In addition, colorectal cancer (CRC) and brain tumours are two dangerous types of cancer, affecting both men and women worldwide \cite{sirinukunwattana2016locality
}. Moreover, there are few studies to estimate the risk factors for knee diseases, which are considered to be the main contributor to elderly disability. This has encouraged the production and application of various artificial intelligence (AI) techniques in medical diagnosis. 

Deep learning (DL) algorithms have played a significant role in this field by automatically extracting the feature engineering process from a wide range of complex problems with deeper layers, thus helping radiologists make faster and more accurate diagnoses \cite{ker2017deep
}. The convolutional neural network (CNN) is one of the most effective deep learning algorithms, which has enabled remarkable advances in the field of medical imaging \cite{lecun2015deep}. The capability of CNNs to detect local elements in an image in a hierarchical way is what gives them their effectiveness, as the high-level layers of CNN are capable of learning more complicated features, whereas the low-level layers are intended to encode general representations for the majority of vision tasks. The most common method to train a CNN architecture is to transfer the learnt knowledge from a previously trained network that completed one task to a new task \cite{pan2010survey}. Because this solution is efficient and simple to implement without the need for a large annotated dataset for training, many researchers prefer to use it, particularly in medical imaging. However, in the healthcare domain, building a robust classification model for datasets with an imbalanced distribution within classes can be a difficult task. Class decomposition (CD), also called Divide-and-Conquer learning, is a supervised approach that enhances boundary learning between given classes by clustering them prior to network training through transfer learning, particularly beneficial for addressing accuracy issues in datasets with imbalanced class distributions \cite{abbas2020detrac, abbas2021classification}.

Curriculum learning (CL) is a way to train a machine learning model that can be used to speed up the training process and improve the performance of the final task by changing the behaviour and learning method of the model \cite{elman1993learning}. The idea behind this particular class of machine learning models draws inspiration from the way humans typically learn. Just as we acquire skills and knowledge in a sequential manner, starting with the basics and gradually progressing to more complex topics, this approach to machine learning follows a similar pattern \cite{bengio2009curriculum}. 
Likewise, CL starts by training the model on simpler examples first, and once the model has started to learn those tasks, we can gradually introduce more complexity into the training data (traditional CL). By doing this, we give the model the opportunity to learn from the simpler features first and converge on the harder ones later. On the other hand, the opposite strategy of CL, known as the ``anti CL'', is to start training with the difficult examples first and then move towards the easier ones \cite{mermer2017scalable}. 
CL involves training a model on simpler tasks before gradually introducing more complex ones. This approach, also known as traditional CL, allows the model to learn from the simpler features first and then converge on the more challenging ones. In contrast, the ``anti-CL'' strategy flips this approach on its head by starting with the difficult examples and then moving towards the easier ones. While this may seem counter-intuitive, it can be useful in certain scenarios where the model needs to quickly adapt to new and challenging tasks \cite{mermer2017scalable}.
The application of the CL strategy involves two factors: a) the ``scoring function'' or the training scheduler, which determines the curriculum strategy to update the training process, and b) the ``pacing/speed functions'' define how quickly to introduce more challenging examples to the model during training.

Medical datasets often exhibit irregularities in data distribution and significant overlap between classes, posing substantial challenges to conventional classification methods. State-of-the-art models still struggle to learn precise class boundaries, leading to reduced performance and reliability. Therefore, there is a pressing need for innovative training approaches that can adapt to these complexities and enhance the robustness of classification systems.

This paper proposes a novel training approach called Curriculum Learning based on Oscillating Granularity of Class Decomposed (\emph{CLOG-CD}) classification. \emph{CLOG-CD} is designed to cope with any irregularities in the data distribution, which is a very common problem in the medical imaging domain, where different classes overlap widely. Class decomposition provides an effective solution to address such a challenging problem by simplifying the learning of class boundaries between the original classes of a dataset. However, the performance of classification models is sensitive to the granularity of the class decomposition. In our work, to improve the robustness in dealing with data irregularities problems, \emph{CLOG-CD} is proposed to control the way CL is learning based on the granularity of class decomposition. For example, \emph{CLOG-CD} can start the training at a high level of granularity (where a better understanding of class boundaries can be achieved) and then move towards a low level of granularity (where the boundaries between classes are more complex to be defined) until reaching the original classes of the dataset. In this way, \emph{CLOG-CD} can gradually learn more specialised features associated with the granularity level of class decomposition. The higher the granularity, the easier it becomes for \emph{CLOG-CD} to discover salient features. 

To the best of our knowledge, this is the first attempt to guide the CL using class decomposition to improve the transferability of features and hence increase the generalisability of deep learning when coping with complex image datasets. The main contributions of this work are summarised as follows:

\begin{itemize} 
\item \emph{CLOG-CD} pioneers the integration of class decomposition into curriculum learning for classification tasks, representing a groundbreaking first in applying this powerful technique within curriculum learning. 


\item By employing anti-curriculum learning with oscillating granularity, \emph{CLOG-CD} empowers models to progressively learn features at multiple levels of decomposition, significantly boosting accuracy and enhancing feature transferability.


 \item Specifically designed to address irregularities in medical image datasets, \emph{CLOG-CD} simplifies complex within-class patterns early in the learning process. This strategic approach enables the model to progressively learn relevant features, substantially improving robustness and generalisation.


\item We conclusively demonstrate the robustness of the proposed method through extensive experiments on four diverse medical image datasets, employing various oscillation steps.

\end{itemize}

The paper is organised as follows: Section 2 reviews the previous works of CL in medical image classification. Section 3 discusses the theoretical analysis of the CL. Section 4 describes our proposed method. Section 5 illustrates the results of our proposed method. Section 6 discusses and concludes our work.


\section{Curriculum learning}
\label{relatedwork}
This section provides an overview of the relevant work on curriculum learning strategies in medical image applications.
In this section, we provide an overview of the relevant works on curriculum learning strategies and multi-pulse variations in medical image applications.
CL has been used in several machine learning tasks to improve the performance of the model, including computer vision \cite{guo2018curriculumnet}, natural language processing \cite{platanios2019competence}
. In the medical image domain, CL methods have proven to be effective in different tasks, especially for the classification task \cite{xie2021survey}. For example, in \cite{lotter2017multi}, the authors built two separate models to detect breast cancer, where they used multi-scaled CNN models based on image patches for segmentation masks of lesions in mammograms. Then the extracted features were fed into a new model based on the whole image level to make the final decision. In \cite{luo2022deep}, the authors proposed a CL model to classify the digital mammogram samples into three classes. The model was trained based on a VGG16 pre-trained network with 5-fold cross-validation from the easy (binary classification) task into the hard (three classes) task. In \cite{jesson2017cased}, the authors introduced a model called CASED for the detection of pulmonary nodules based on an imbalanced CT image dataset. The difficulty of the curriculum was represented by the size of the input nodules, where the model first learnt to distinguish nodules from their immediate surroundings, and then more global context was gradually added. In \cite{park2019curriculum,cho2021optimal}, two CL strategies were developed to detect various diseases in CXR images. Firstly, the regions of interest (ROIs) were identified based on patch images, and then the Resnet-50 pre-trained network was fine-tuned using the whole images. The results showed that CL strategy performance was better than the training with the baseline ResNet-50. In \cite{wei2021learn}, a CL model was proposed for the classification of colorectal polyp images. The model was trained based on Resnet-18 with four different image combinations, starting at an easy level and gradually increasing in complexity.
In \cite{jimenez2019medical}, a CL model was proposed for the classification of proximal femur fracture into two cases, where the authors proved that starting the training model in ascending-descending order had achieved better performance for fracture classification.


\section{CLOG-CD Model}
\label{CLOG-CD}
\begin{figure*}[h!]
  \centering
    \includegraphics[scale=0.35]{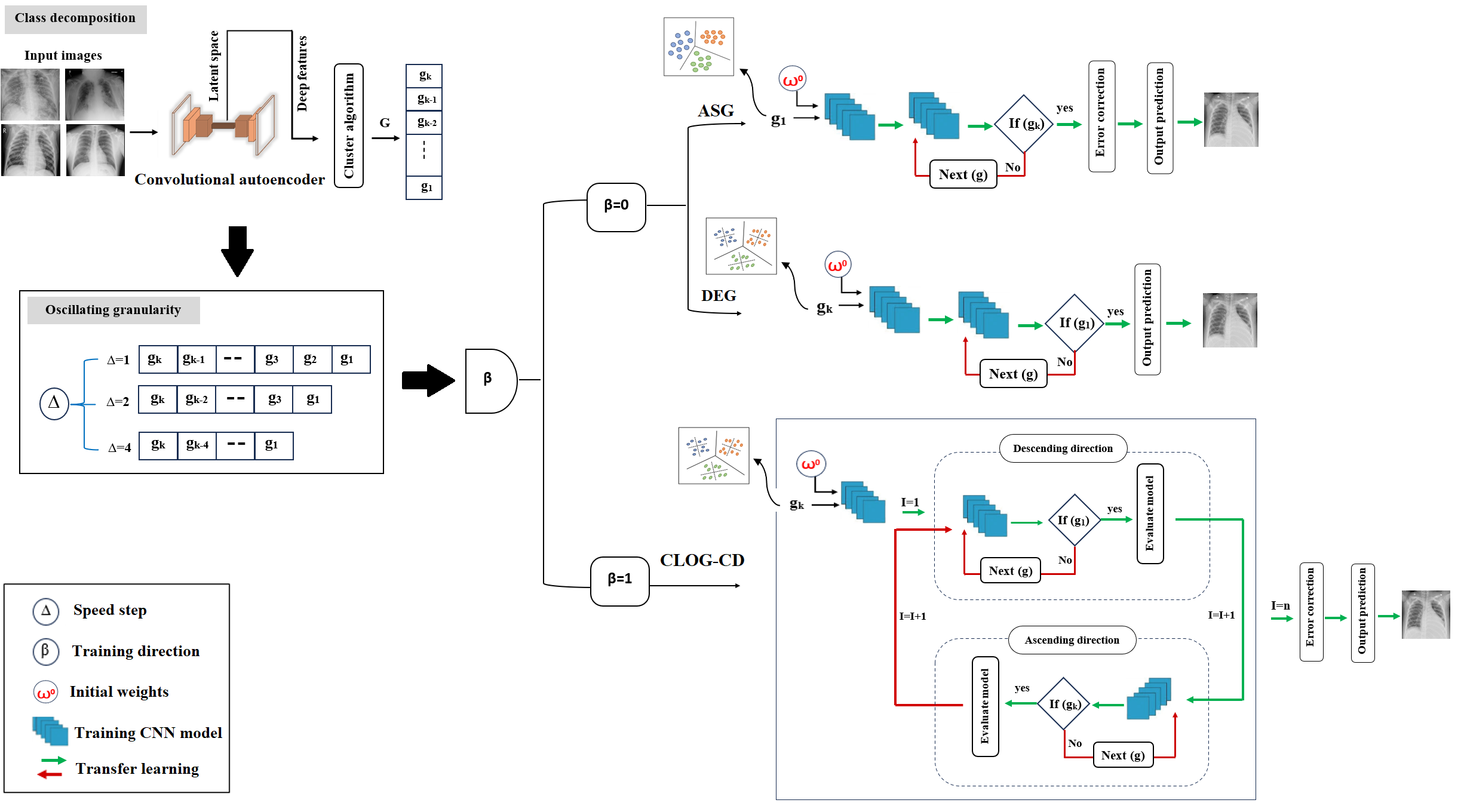} 
  \caption{The framework of the \emph{CLOG-CD} model. The dataset is decomposed into sequential granularity levels with $k$ components based on different speeds. The model fine-tunes initial weights $(w^{0})$, starting with the original classes of the dataset $(g_{1})$ in the (ASG) model. While, in (DEG) and \emph{CLOG-CD} with different speeds, training starts at the maximum granularity level $(g_{k})$, where the dataset was decomposed into the maximum number of sub-classes.}
\label{Method}
\end{figure*}

This section describes in sufficient detail our proposed training method, Curriculum Learning based on Oscillating Granularity of Class Decomposed images (\emph{CLOG-CD}), see Fig. \ref{Method}. We also discuss the impact of CL based on the class decomposition and formalise the method.

\subsection{Class Decomposition}

First, we extracted deep local features from the data augmentation (AUG) technique using a convolutional autoencoder (CAE). CAE contains two CNN blocks: the encoder, which is used for compressing the input image into a lower-dimensional latent representation, and the decoder which is responsible for reconstructing the original image as it was. It uses the convolution operator to extract features from the data by scanning the entire image with squared convolutional filters.
Assume that the training dataset D=$\left \{ \left ( x_{i},y_{i} \right ) \right.\left. ,i=1,...N \right \}$, where $x_{i}\epsilon X$ is the training set,  $y_{i}\epsilon Y$ are the corresponding labels, and N is the total number of examples. Then the generated activation maps of the encoder process for the input $x$ can be defined as:

\begin{equation}
   A^{d} = \alpha \left ( x \times {W}^{d} + {b}^{d}\right ),
\end{equation}
where \( \mathbf{\alpha} \) is an activation function, $W$ is the weights of the square convolution filter, and $b^{d}$ is the bias for d-th activation maps. 

\begin{figure*}[h!]
  \centering
    \includegraphics[scale=0.5]{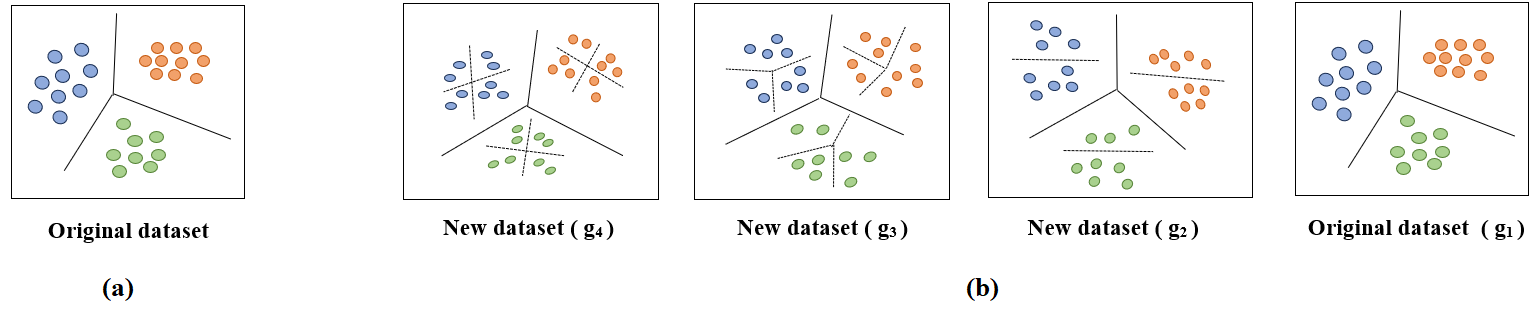} 
  \caption{An example of the granularity of the class decomposition method: a) the original class, b) the newly generated datasets after applying the granularity of class decomposition (e.g. $k$=4).}
 \label{granularity}
\end{figure*}

Second, the vector derived from the latent space is decomposed into a sequence of granularity levels using the $k$-means algorithm \cite{wu2008top}. Let \textbf{G} represent this sequence of granularity decomposition in descending order. For example, as shown in Fig. \ref{granularity}, if $k$=4, this means we generate a dataset where each class is divided into four sub-classes, which is denoted as $g_{4}$. Additionally, we generate datasets $g_{3}$ and $g_{2}$ where each class is divided into 3 and 2 sub-classes, respectively, and $g_{1}$ corresponds to the original dataset without any decomposition. Hence the granularity of decomposition based on descending order is represented as the sequence: \begin{math} \textbf{G}=\left\{ g_{k},g_{k-1},\cdots ,g_{1}\right\} \end{math}.
where each $g_{1}$ represents a dataset with the classes divided into i sub-classes, $g_{1}$ represents the original dataset without decomposition, and \textbf{$g_{k}$} corresponds to the dataset formed after applying the maximum number of $k$ decomposition components.
  

In our experiment, the decomposition process was applied using $k$-means cluster algorithm ($k$=5), where each data point in the class $c$ is picked by the closest cluster centroid $c_{j}$ according to the squared euclidean distance (SED):


 \begin{equation}
     SED= \sum _{j=1}^{k} \sum _{i=1}^{n}\parallel x_{i}^{ \left( j \right) }-c_{j}\parallel^{2}, 
    \end{equation}

In the decomposition process, each sub-class is given a new label associated with its original class and treated as an individual new class. Finally, those sub-classes are then recombined after training to compute the error correction of the final prediction and obtain the classification output.





\subsection{Oscillating Granularity}
\label{prposedmethod}
Consider the learning task that aims to improve the training of the prediction function $f_{\theta} : X\rightarrow Y$, where $\theta$ is the trainable parameters, over the dataset D, using knowledge acquired by learning a series of n functions: $(f_{\theta_{1} },...,f_{\theta_{n} })$.

Then the ``scoring function (S)''  based on descending order can be written as: 
\begin{equation}  
     S(x_{i},y_{i}) > S(x_{j},y_{j}),   \forall \, S: X \to R,
\end{equation}

where the data point $S(x_{i},y_{i})$ is more difficult than $S(x_{j},y_{j})$. In the training process, we used mini-batches (MB) with stochastic gradient descent (SGD). The collection of mini-batches is denoted as ${MB=\{B_{1}, B_{2},...B_{M}}\}$, where MB $\subseteq$ X and $M$ is the number of subsets in the training set. Let $X\acute{_{i}}$ be a subset of $B_{i}$, then the pacing function ($P$) can be defined as:

\begin{equation}
    P_{\theta}(i)= \left| X\acute{_{i}} \right|
\end{equation}

where ${X\acute{_{i}}=\{X\acute{_{1}},X\acute{_{2}},....,X\acute{_{M}}}\}$ represents the subset of samples within the mini-batch $B_{i}$ when sorted by the scoring function ($S$).

In our work, we explored different speed steps to evaluate the effectiveness of our proposed method. Let $\bigtriangleup$ represents the size of the oscillation step, $\bigtriangleup= (1,2,...,n)$. This step determines when the CL strategy is updated to progress to the next level ($l$) in \textbf{G}, and let $\beta$ refers to the direction of the training process and can be defined as follows:   

\begin{eqnarray}
    \beta=\begin{cases}
    0: & \text{Descending granularity}.\\
    1: & \text{Ascending + Descending granularity}.
  \end{cases}
\end{eqnarray}




   
We first investigated the effectiveness of the \emph{CLOG-CD} model without oscillation changing, and in one direction (i.e. $\bigtriangleup=1$, $\beta=0$) based on ascending order, we called this process ``\emph{CLOG-CD} (ASG)''. Where the initial model starts training from the original classes (i.e. $g_{1}$), then the converged learned weights are transformed into the subsequent curriculum levels (i.e. $g_{2}$), and gradually progressing until we reach the maximum level (i.e. $g_{5}$). Second, we evaluated the model based on descending order, we called this process ``\emph{CLOG-CD} (DEG)'', where $\bigtriangleup=1$, $\beta=0$. Here, the model starts training from the hard level (i.e. $g_{5}$), and then the learned weights are transformed into the subsequent curriculum level below ($g_{4}$) and gradually until reaching the original classes ($g_{1}$). At the end of these processes, we computed the overall classification performance on the test set.


We then repeated the \emph{CLOG-CD} (DEG) process in the opposite direction over $I$ times, I=20 for ResNet-50 and I=10 for DenseNet-121. The model is trained starting with the descending granularity order, and then returning to the ascending granularity order. We called this process ``\emph{CLOG-CD} $(\bigtriangleup=1)$'', where $\bigtriangleup = 1$ and $\beta= 1$. Finally, we evaluated the performance of \emph{CLOG-CD} based on two different oscillation steps; we defined them as ``\emph{CLOG-CD} $(\bigtriangleup=2)$''  and ``\emph{CLOG-CD} $(\bigtriangleup=4)$''. In the \emph{CLOG-CD} $(\bigtriangleup=2)$ process, the initial model starts training at the maximum granularity ($g_{5}$) and moves towards the level ($g_{3}$) and finally to the lowest level ($g_{1}$). While in \emph{CLOG-CD} $(\bigtriangleup=4)$ process, the initial model starts training at the maximum granularity ($g_{5}$) and progresses to the lowest level ($g_{1}$) before returning to the highest granularity. The procedural steps of the \emph{CLOG-CD} model are summarised in Algorithm \ref{Algorithm1}.





\begin{algorithm}[h!]
\setstretch{0.94}
\caption{\emph{\emph{CLOG-CD} Model}}
\label{Algorithm1}
\textbf{Input:} Unlabelled samples, labelled dataset, $\bigtriangleup$: oscillation step, $\beta$: training direction, $k$: cluster component, $I$: number of iterations.

\textbf{Output:} G: new datasets generated by using class class decomposition method, prediction output.

\textbf{Granularity of Class Decomposition:} \\
      \quad \quad Use CAE for training unlabelled samples.\\
    \quad \quad Extract features from the latent representation.\\
  \quad \quad  Use $k$-means to generate \textbf{G} in descending order.\\
   \quad \quad  \begin{math} \textbf{G}=\left\{ g_{k},g_{k-1},\cdots ,g_{1}\right\} \end{math}\\
\textbf{Training \emph{CLOG-CD} on one direction:}\\
\quad \quad Training with ($\bigtriangleup=1$, $\beta=0$).  \\
\If{process = \emph{CLOG-CD}(ASG)}{
    Arrange \textbf{G} in ascending order.}

\For{i in \textbf{G}}{
    \If{i = 1}{
        $w' \gets$ Initial training (pre-trained network, $\textbf{G}[i]$).}
    $W$: the best learned weights.\\
    Training the model ($W$, $\textbf{G}[i]$).}
Evaluate the performance.
         
\textbf{Training \emph{CLOG-CD} based on both directions:} \\
  \quad  Training with ($\bigtriangleup=[1, 2, 4]$, $\beta=1$).\\
   \quad  $w' \gets$ initial training (pre-trained network, $g_{k}$)

$I = 0$ \\
\While{$I < n$} {
    $I = I + 1$ \\
    \For{$i$ in $\textbf{G}$} {    
        $W$: Transfer $w'$ only at I=1.\\
        $\textbf{G} \gets$  Descending order.\\        Training the model ($W$, \textbf{G}[i]). \\
        $W$: the best learned weights. \\  }
    Evaluate on the test set.\\
    $I = I + 1$ \\
    \For{$i$ in $\textbf{G}$} {
        $\textbf{G} \gets$  Ascending order. \\
        Training the model ($W$, \textbf{G}[i]). \\
        $W$: the best learned weights. }
Evaluate on the test set. }
Select the best performance among $I_n$.
\end{algorithm}

\subsection{Performance Evaluation}
For performance evaluation, we adopted several metrics of the multi-classification problem, including precision, recall, and F1 score \cite{sokolova2009systematic}, see Section \ref{sec:Evaluation}. In addition, we computed the 95\% confidence interval (CI) over I iterations for each dataset to provide a robust evaluation of our model’s performance \cite{efron1987better}.

\section{Experimental setup and results}
\label{results}
This section describes the datasets used to investigate the effectiveness of our approach \emph{CLOG-CD}, and discusses the experimental results. 

\subsection{Datasets description}
In this work, we used four different datasets: CXR, brain
tumour, digital knee x-ray, and CRC datasets. These datasets have been selected due to the imbalanced problem between the classes of each dataset. 
The CXR dataset was collected by the researchers of Qatar University \cite{chowdhury2020can, rahman2021exploring}, and comprised of four classes, see Table \ref{distribution_CXR}. Interventional Radiology and the Italian Society of Medical Radiology collected the images in COVID-19 database. All images are in PNG format with a resolution of $299 \times 299$ pixels; see Fig. \ref{fig:SampleLung}. For the brain tumour dataset, we used 3064 images from Nanfang and General Hospitals, Tianjin Medical University, China, with three types of acquired brain tumours \cite{badvza2020classification}, see Table \ref{distribution_brain}. All the images are $400 \times 400$ pixels and in PNG format, see Fig. \ref{Samplebrain}. The digital knee x-ray consists of 1,650 knee MRI images and was obtained from well reputed hospitals and diagnostic centres \cite{gornale2020digital} using PROTEC PRS 500E x-ray machine with the help of two medical experts, 8-bit gray-scale graphics were used in the original images with PNG format, see Fig. \ref{Sampleknee}. In our work, we used the images contained in sub-folder “MedicalExpert-I” which labelled into five classes, see Table \ref{distribution_knee}. For the CRC dataset, we used the dataset "CRC-VAL-HE-7K" from UMM (University Medical Center Mannheim, Heidelberg University, Mannheim, Germany) \cite{macenko2009method}. The dataset contains nine imbalanced classes, see Table \ref{distribution_CRC}. The images are $224 \times 224$ pixels with TIF format, see Fig. \ref{SampleCRC}.

We evaluated the performance of our proposed method with ResNet-50 \cite{he2016deep} and DenseNet-121 \cite{huang2017densely}, and the models were trained based on a deep-tuning strategy. All images were resized to $224 \times 224$ pixels to be compatible with the pre-trained networks, and we used the bilinear interpolation technique for the resizing process, which is commonly used in image processing to maintain image quality, critical features and minimise artifacts. Then, we randomly divided the dataset into three sets: 70\% for the training set, 20\% for the validation set, and 10\% for the test set, which was used for evaluating the performance. 

All experiments were carried out using the Python programming, Keras library on a virtual machine with a processor Intel(R) Core(TM) 32 Duo @ 2.40 GHz, NVIDIA Quadra P8000GPU, and 64.00 GB for RAM capacity. To optimize the model during training, we used a cross-entropy loss function with mini-batch stochastic gradient descent (mSGD) and based on deep-tuning mode for 50 epochs with a mini-batch size of 50.




\begin{table}[h!]
\begin{center} 
\caption{Experimental hyper-parameter settings for each dataset. }
\label{Hyperparme}
\small
\tabcolsep=0.10cm
\resizebox{8.7cm}{!}{
\renewcommand{\arraystretch}{1.2}
\begin{tabular}{c|cc|cc}
\hline
&\multicolumn{2}{c|}{ResNet-50}& \multicolumn{2}{c}{DenseNet-121}\\
Dataset  & Learning rate  &  Learning rate-decay  & Learning rate  &  Learning rate-decay \\
\hline
CXR   & 0.001 &  0.85 every 10 epochs& 0.001 & 0.80 every 15 epochs \\
Brain tumour  & 0.0001  &  0.9 every 10 epochs& 0.001&  0.80 every 10 epochs     \\
Knee x-ray  &  0.001 &  0.90 every 15 epochs& 0.0001 & 0.85 every 10 epochs  \\
CRC    & 0.0001  &  0.95 every 15 epochs& 0.001 &  0.90 every 15 epochs     \\
\hline
\end{tabular}
}
\end{center} 
\end{table}

\subsection{Class decomposition of \emph{CLOG-CD}}
We construct a convolutional autoencoder (CAE) with two convolutional layers, 3 $\times$ 3 kernel size, and ReLU as an activation function. For training in the CRC and digital knee x-rays, the number of filters in the first layer was set at 32, and the number of filters in the second layer was set at 16. For the CXR and brain tumour datasets; the number of filters in the first and second layers was set to 16 and 8, respectively. Adam optimizer 
was used to train the models, and the learning rate was set to 0.001 with 50 epochs number and 50 for a mini-batch size. The extracted features from the latent representation were then fed into the $k$-means clustering algorithm to generate granularity clusters of decomposition. Here, we applied a class decomposition process with $k$=5, resulting in four new datasets with new sub-classes related to the original class label of each dataset. Also, we experiment with the classification performance based on the original classes without applying CL strategy.\\

\subsection{Evaluation of \emph{CLOG-CD}}
In this subsection, we evaluated the performance of our proposed model, \emph{CLOG-CD}, using four different medical image datasets. The evaluation includes comparing the model's performance through three different speed steps and examining different training strategies, including an ascending-descending CL strategy \emph{CLOG-CD} (ASG), anti-CL strategy \emph{CLOG-CD} (DEG), and training a baseline model using two pre-trained networks (ResNet50 and DenseNet-121). Additionally, we evaluated the effectiveness of the proposed method before and after applying data augmentation techniques on the training set. 
ResNet-50 is a deep CNN with 50 layers, including 48 convolutional layers, 3 $\times$ 3 Max-Pooling layer, and 1 $\times$ 1 Average-Pooling layer followed by the classification layers. While DenseNet-121 consists of 120 convolutional layers followed by a fully connected layer. It consists of multiple dense blocks, each block includes 1$\times$1 convolution and 2 $\times$ 2 average pooling operations, which help reduce the number of feature maps and maintain computational efficiency. 
To prevent the overfitting, we used the regularization technique-L2 with value 0.001 for CXR, brain tumour, and digital knee x-ray, and 0.0001 for the CRC dataset. The parameter settings for training each dataset are reported in Table \ref{Hyperparme}, except the the last fully connected layer was set to 0.01.


\begin{itemize}

    \item \textbf{Classification performance on CXR dataset}\\

We investigate the classification performance of the \emph{CLOG-CD} on the 2,119 CXR test images after applying the AUG techniques, see Table \ref{distribution_CXR}. The last fully connected layer was adapted to the sub-classes of each new dataset. First, we evaluated the performance with the baseline classifier (ResNet-50) on classification four classes. The obtained results were (91.65\%, 93.82\%, 91.99\%, and 92.90\%) for ACC, PR, RE, and F1, respectively.
Next, we compared the results with the classification performance of \emph{CLOG-CD} (ASG) process, where the model was trained with a traditional CL strategy in a single direction from ascending to descending order. The model achieved overall ACC of 89.52\%, with PR, RE, and F1 score of 90.53\%, 89.06\%, and 89.79\%, respectively. For the \emph{CLOG-CD} (DEG) process, where the model was trained based on anti-CL strategy in a single direction, the results improved to 93.16\% for ACC, 94.84\% for PR, 94.18\% for RE, and 94.51\% for F1. Second, we evaluated the classification performance of \emph{CLOG-CD} based on three different curriculum accelerations in both directions over 20 iterations, where the model started training in descending order and then returned to ascending order. The outcomes from \emph{CLOG-CD} ($\bigtriangleup=1$) process achieved the highest performance with 96.08\% for ACC, 97.16\% for PR, 96.71\% for RE, and 96.94\% for F1. In the \emph{CLOG-CD} ($\bigtriangleup=2$) process, the obtained results were 95.66\%, 96.63\%, 96.30\%, and 96.46\% for ACC, PR, RE, and F1, respectively. While in \emph{CLOG-CD} ($\bigtriangleup=4$), the results showed a decrease in the performance with values of 94.86\% for ACC, 95.72\% for PR, 95.16\% for RE, and 95.44\% for F1. The results are reported in the first row in Table \ref{result_Resnet} with AUG techniques. Additionally, the second part of Table \ref{result_Resnet} shows the results for each training strategy without AUG processes. The overall ACC for the baseline (ResNet-50) reached 89.76\%, while the \emph{CLOG-CD} (ASG) process achieved 88.44\%, and the \emph{CLOG-CD} (DEG) process reached 89.75\%. On the other hand, the \emph{CLOG-CD} ($\bigtriangleup=1$) process achieved the highest ACC at 93.58\% compared to \emph{CLOG-CD} ($\bigtriangleup=2$) and \emph{CLOG-CD} ($\bigtriangleup=4$) processes which achieved 88.01\% and 87.54\%, respectively. 

Likewise, we compared the baseline DenseNet-121 with other training strategies using CL, see Table \ref{result_DenseNet}. The observed results demonstrated that the classification performance of the \emph{CLOG-CD} (DEG) achieved a higher accuracy (91.69\%) compared to the baseline and \emph{CLOG-CD} (ASG), which achieved 88.72\% and 86.74\%, respectively. While the results from the \emph{CLOG-CD} ($\bigtriangleup=1$) process achieved the highest accuracy (94.86\%) compared to the \emph{CLOG-CD} ($\bigtriangleup=2$) and \emph{CLOG-CD} ($\bigtriangleup=4$), which achieved 93.44\% and 91.27\%, respectively. Table \ref{CI} illustrates the CI at 95\% over 20 iterations with ResNet-50 and 10 iterations with DenseNet-121.

Fig. \ref{Time_CXR} and Fig. \ref{Time_lung_dense} show the fitted curves obtained by different oscillation steps with ResNet-50 and DenseNet-121, respectively. Where the black curve refers to \emph{CLOG-CD} ($\bigtriangleup=1$) process, which achieved the highest classification performance, while the red and green curves represent the \emph{CLOG-CD} ($\bigtriangleup=2$) and \emph{CLOG-CD} ($\bigtriangleup=4$) processes, respectively. Furthermore, Fig. \ref{cm_CXR} and Fig. \ref{cm_cxy_denese} illustrate the confusion matrices for each model of CXR dataset.\\

\begin{figure*}[h!]
   \centering
      \includegraphics[scale=0.53]{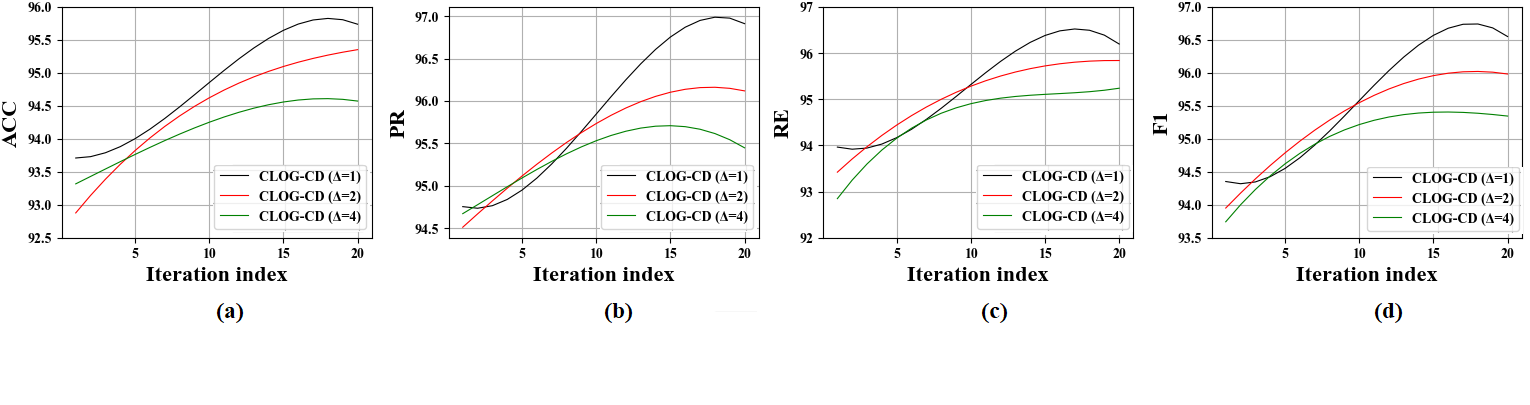} 
     \caption{The fitted curve at degree 3 of CXR dataset over 20 iterations obtained by: a) ACC, b) PR, c) RE, and d) F1, with ResNet-50. }
    \label{Time_CXR}
\end{figure*}

\begin{figure*}[h!]
   \centering
      \includegraphics[scale=0.22]{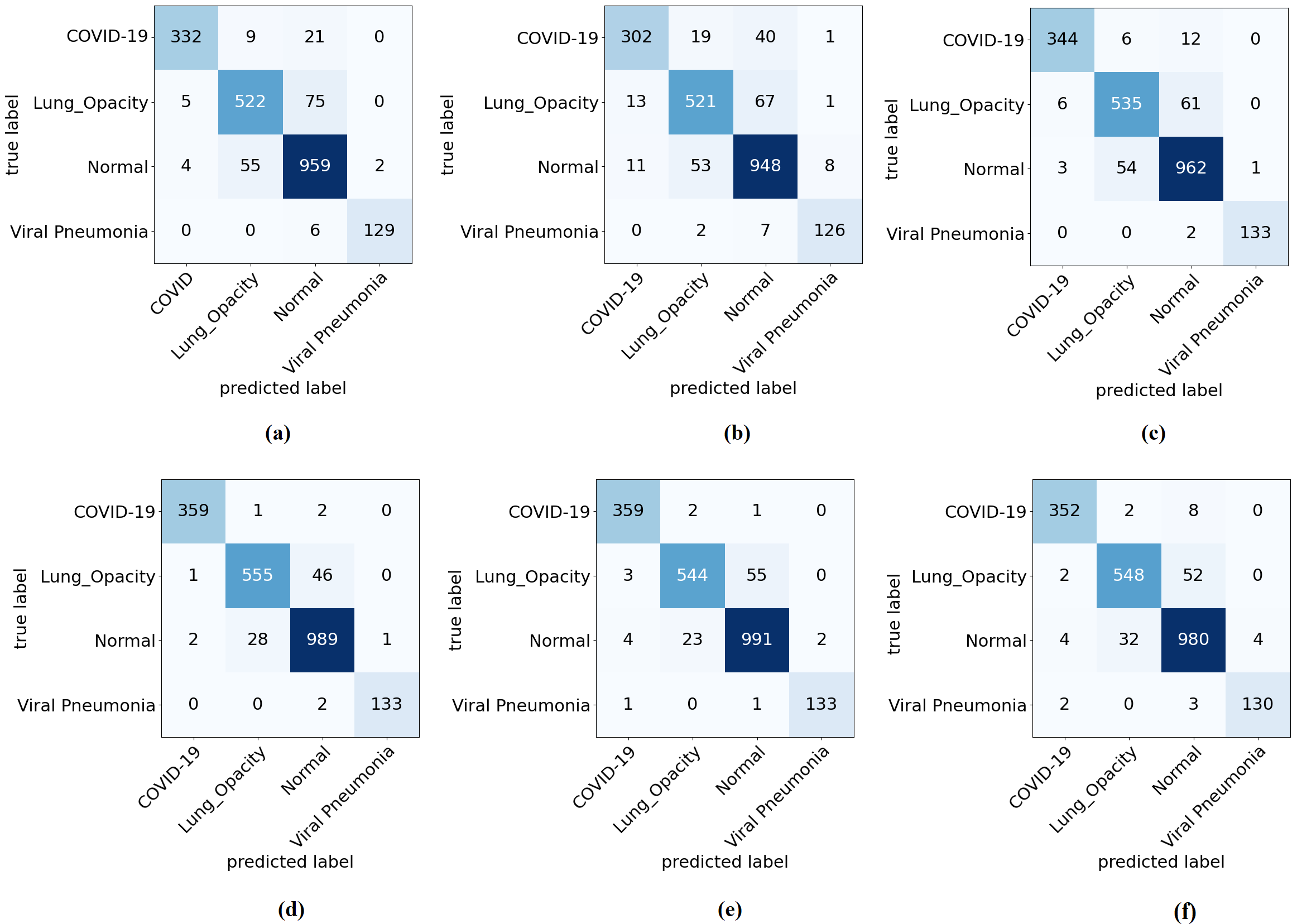} 
     \caption{The confusion matrix results of CXR dataset obtained by: a) ResNet-50 baseline, b) \emph{CLOG-CD} (ASG), C) \emph{CLOG-CD} (DEG), d) \emph{CLOG-CD} ($\bigtriangleup=1$), e) \emph{CLOG-CD} ($\bigtriangleup=2$), and f) \emph{CLOG-CD} ($\bigtriangleup=4$).}
    \label{cm_CXR}
\end{figure*}



\item \textbf{Classification performance on brain tumour dataset}\\

For further investigation, we compared the performance of \emph{CLOG-CD} with the baseline for classification brain tumour test sets into three classes. As shown in Table \ref{result_Resnet}, the overall classification performance of the baseline ResNet-50 reached at 90.89\%, while the \emph{CLOG-CD} (ASG) process recorded 77.56\% for ACC. In contrast, the \emph{CLOG-CD} (DEG) process achieved a higher accuracy of 92.20\%. For the \emph{CLOG-CD} ($\bigtriangleup=1$) with AUG techniques, the results recorded a slight increase in accuracy at 96.91\%, compared to \emph{CLOG-CD} ($\bigtriangleup=2$), and \emph{CLOG-CD} ($\bigtriangleup=4$) which achieved 96.75\% and 95.12\%, respectively. Without AUG, a significant improvement was observed, with an ACC of 90.73\% for \emph{CLOG-CD} ($\bigtriangleup=1$), compared to 87.97\% for \emph{CLOG-CD} ($\bigtriangleup=2$) and 84.88 for \emph{CLOG-CD} ($\bigtriangleup=4$). 

Regarding baseline DenseNet-121, the \emph{CLOG-CD} ($\bigtriangleup=1$) process with AUG achieved the highest performance with 94.63\%, 93.91\%, 93.99\%, and 93.95\% for ACC, PR, RE, F1, respectively. Without AUG, the results were 91.87\%, 90.67\%, 90.53\%, 90.60\% for ACC, PR, RE, and F1, respectively, see Table \ref{result_DenseNet}. Fig. \ref{Time_brain} and Fig. \ref{Time_brain_dense} show the fitted curves obtained with ResNet-50 and DenseNet-121, respectively. As shown, the classification performance of the \emph{CLOG-CD} ($\bigtriangleup=1$) process (the black curve) is higher than other training strategies. Additionally, Fig. \ref{cm_brain} and Fig. \ref{cm_brain_Dense} illustrate the confusion matrices for the brain test set with ResNet-50 and DenseNet-121, respectively. \\

\begin{figure*}[h!]
   \centering
      \includegraphics[scale=0.53]{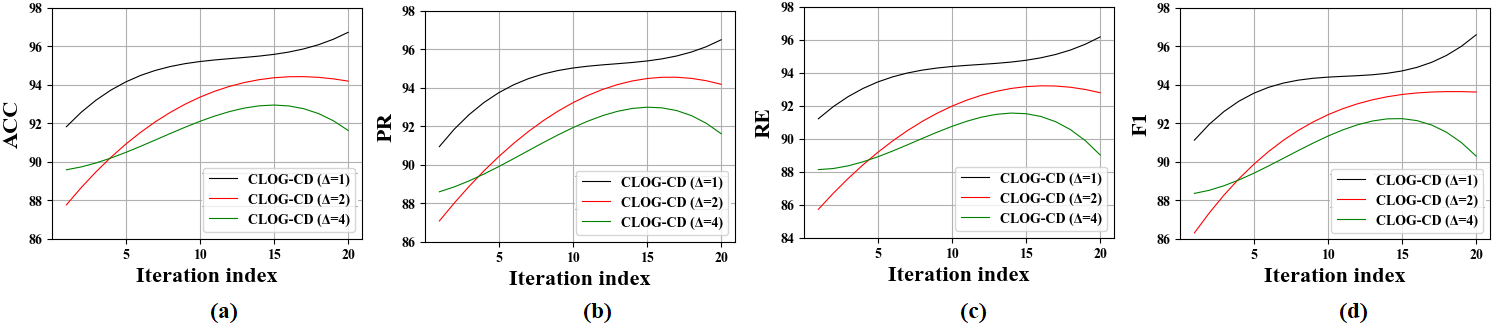} 
     \caption{The fitted curve at degree 3 of brain tumour dataset over 20 iterations obtained by: a) ACC, b) PR, c) RE, and d) F1, with ResNet-50.}
    \label{Time_brain}
\end{figure*}

\begin{figure*}[h!]
   \centering
      \includegraphics[scale=0.2]{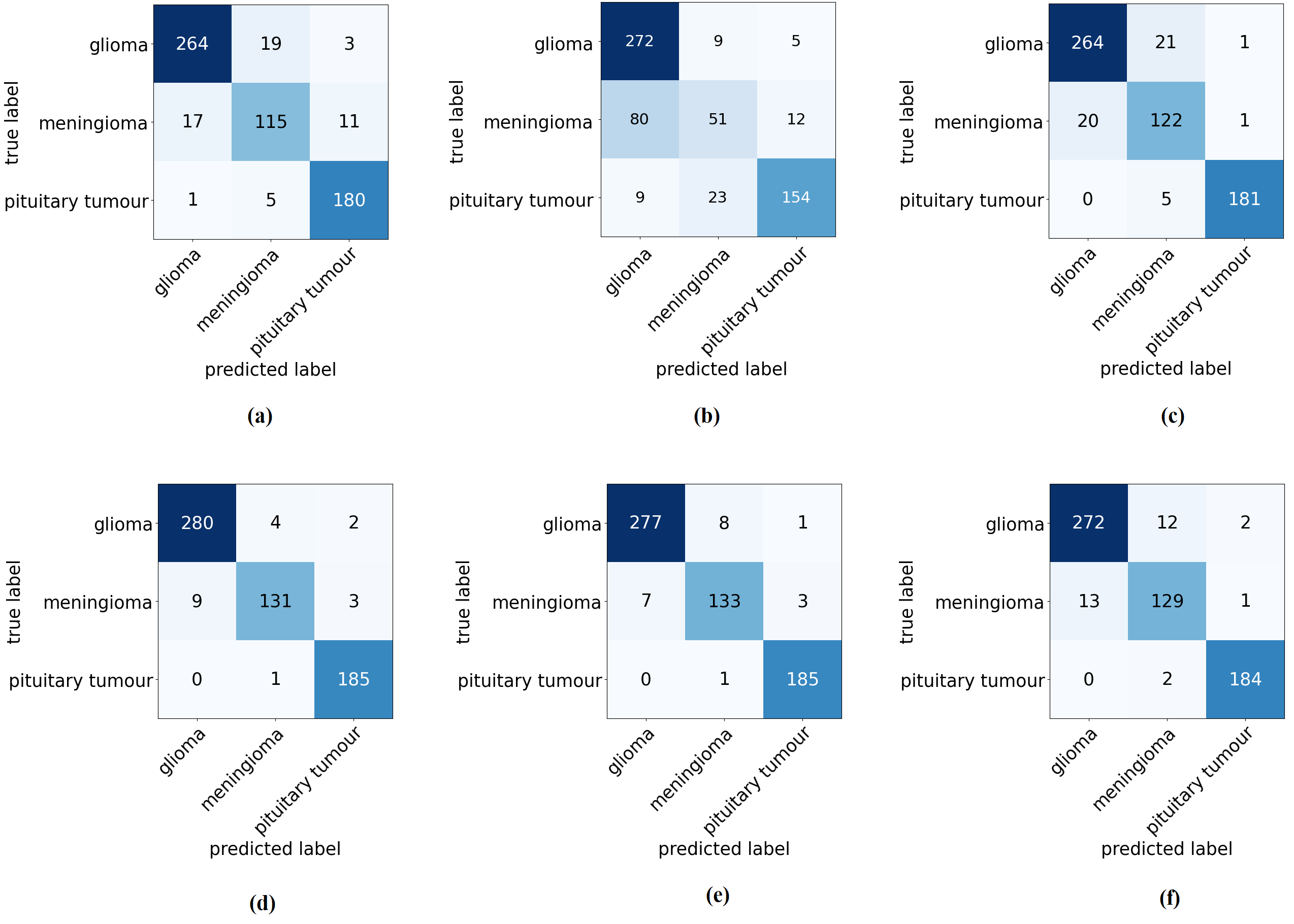} 
    \caption{The confusion matrix results of brain tumour dataset obtained by: a) ResNet-50 baseline, b) \emph{CLOG-CD} (ASG), C) \emph{CLOG-CD} (DEG), d) \emph{CLOG-CD} ($\bigtriangleup=1$), e) \emph{CLOG-CD} ($\bigtriangleup=2$), and f) \emph{CLOG-CD} ($\bigtriangleup=4$).}
    \label{cm_brain}
\end{figure*}


\item \textbf{Classification performance on digital knee dataset}\\

Moreover, we evaluated the performance of the \emph{CLOG-CD} model in classifying the digital knee x-ray test set into five classes, see Table \ref{distribution_knee}. As demonstrated in Table \ref{result_Resnet}, the results obtained using AUG process with the \emph{CLOG-CD} ($\bigtriangleup=1$) model outperformed other training strategies, achieving 79.76\% for ACC, 81.60\% for PR, 78.80\% for RE, and 80.18\% for F1. While the baseline ResNet-50 achieved the lowest performance with 63.69\% for ACC, 61.36\% for PR, 62.72\% for RE, and 62.03\% for F1. Without AUG, the \emph{CLOG-CD} ($\bigtriangleup=1$) model also surpassed the other models, achieving 67.85\%, 69.38\%, 64.40\%, 66.80\% for ACC, PR, RE, and F1, respectively, compared to the baseline which achieved the lowest performance with 35.12\%, 33.84\%, 35.07\%, 34.44 for ACC, PR, RE, F1, respectively. 
Furthermore, from Table \ref{result_DenseNet}, the outcomes from the baseline DenseNet-121 with AUG achieved the lowest classification performance with 60.12\%, 58.29\%, 62.14\%, 60.15\% for ACC, PR, RE, and F1, respectively. In contrast, the \emph{CLOG-CD} ($\bigtriangleup=1$) model showed significant improvement, achieving 76.19\% for ACC, 75.59\% for PR, 76.79\% for RE, and 76.19\% for F1, compared to other training strategies. Similarly, without AUG, the \emph{CLOG-CD} ($\bigtriangleup=1$) model achieved the highest classification performance. Fig. \ref{Time_knee} and Fig. \ref{Time_knee_dense} visualise the fitted curves obtained by different curriculum oscillation processes. Moreover, Fig. \ref{cm_knee} and Fig. \ref{cm_cxy_denese} show the confusion matrices for the knee image test set.\\

\begin{figure*}[h!]
   \centering
     \includegraphics[scale=0.52]{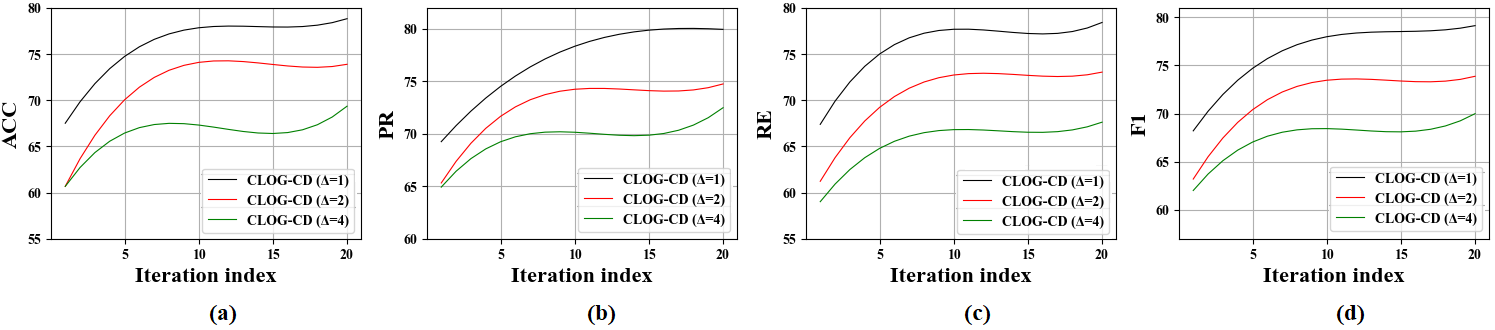} 
     \caption{The fitted curve at degree 3 of digital knee x-ray over 20 iterations obtained by: a) ACC, b) PR, c) RE, and d) F1, with ResNet-50.}
    \label{Time_knee}
\end{figure*}

\begin{figure*}[h!]
  \centering
     \includegraphics[scale=0.24]{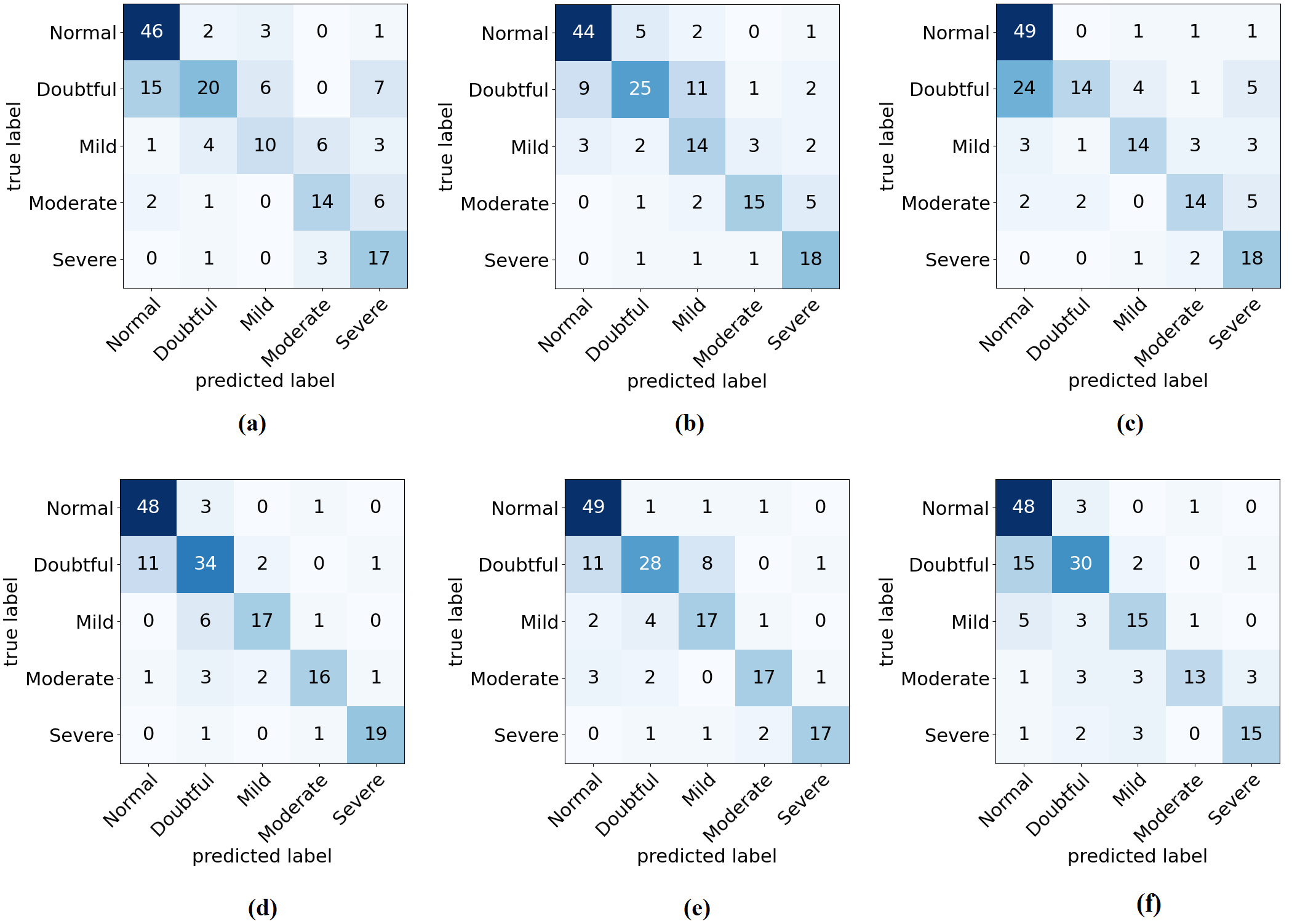} 
     \caption{The confusion matrix results of the digital knee x-ray obtained by: a) ResNet-50 baseline, b) \emph{CLOG-CD} (ASG), C) \emph{CLOG-CD} (DEG), d) \emph{CLOG-CD} ($\bigtriangleup=1$), e) \emph{CLOG-CD} ($\bigtriangleup=2$), and f) \emph{CLOG-CD} ($\bigtriangleup=4$).}
    \label{cm_knee}
\end{figure*}


\item \textbf{Classification performance on CRC dataset}\\

Finally, we investigated the evaluation performance on the test set of the CRC dataset, see Table \ref{distribution_CRC}. As you can see from Table \ref{result_Resnet}, the overall classification performance with AUG from the \emph{CLOG-CD} ($\bigtriangleup=1$) model achieved the highest classification performance with 99.17\% for ACC, 99.12\% for PR, 98.99\% for RE, and 99.06\% for F1. Also achieved a significant improvement without AUG with 88.52\% for ACC, 88.51\% for PR, 88.28\% for RE, 88.39\% for F1 compared to other training models.

For baseline DenseNet-121, the outcomes with AUG process from the \emph{CLOG-CD} ($\bigtriangleup=1$) model also proved its efficiency, achieving 99.45\%, 99.57\%, 99.40\%, and 99.49\% for ACC, PR, RE, and F1, respectively, see Table \ref{result_DenseNet}. Without AUG, \emph{CLOG-CD} ($\bigtriangleup=1$) process also surpassed other training strategies achieving 92.25\% for ACC, 91.13\% for PR, 89.60\% for RE, 90.39\% for F1. Fig. \ref{Time_CRC} and Fig. \ref{Time_CRC_dense} represent the fitted curves obtained by different curriculum oscillation processes, Fig. \ref{cm_CRC} and Fig. \ref{cm_colon_dense} illustrate the confusion matrices obtained by each model for each class in the CRC dataset. 

\end{itemize}

\begin{figure*}[h!]
   \centering
      \includegraphics[scale=0.52]{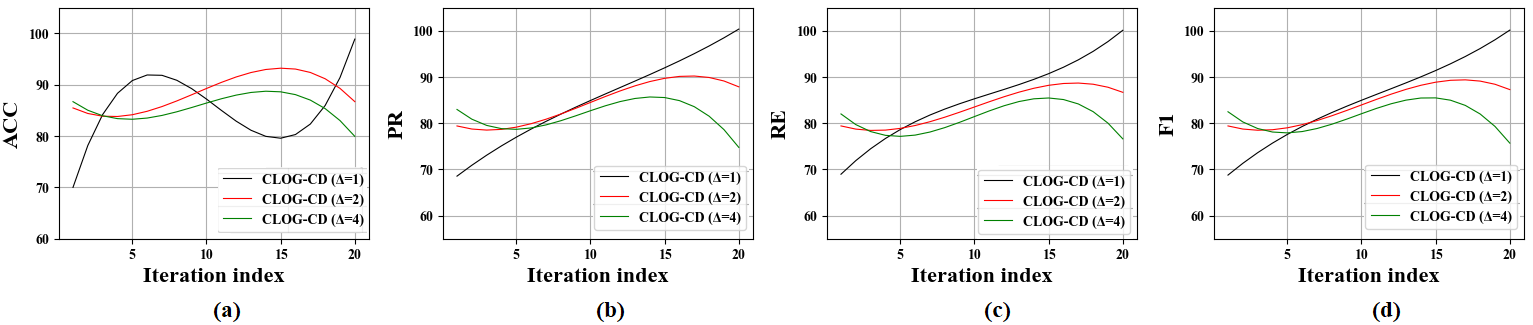} 
     \caption{The fitted curve at degree 3 of CRC dataset over 20 iterations obtained by: a) ACC, b) PR, c) RE, and d) F1, with ResNet-50.}
    \label{Time_CRC}
\end{figure*}

\begin{figure*}[h!]
   \centering
      \includegraphics[scale=0.35]{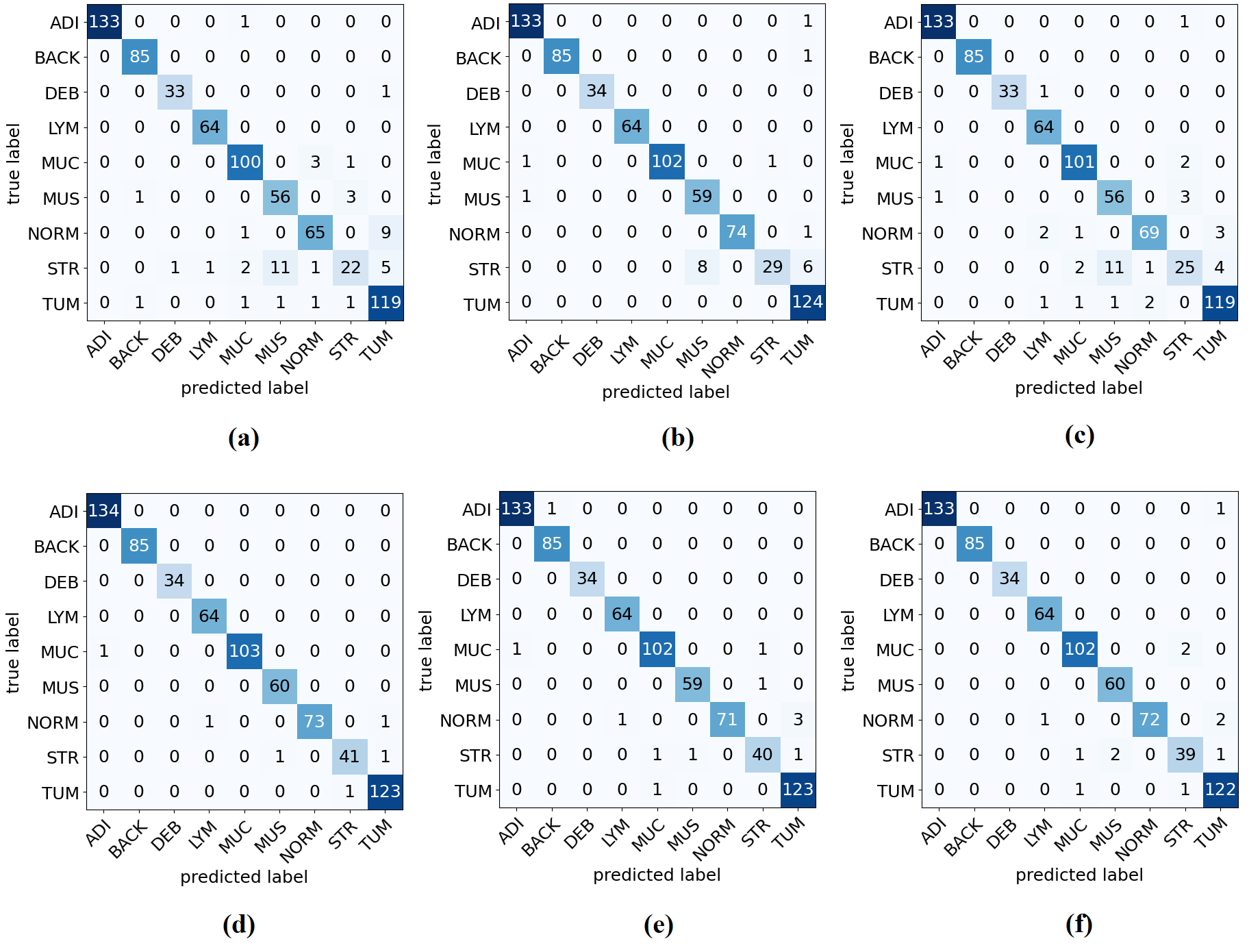} 
     \caption{The confusion matrix results of CRC dataset obtained by: a) ResNet-50 baseline, b) \emph{CLOG-CD} (ASG), C) \emph{CLOG-CD} (DEG), d) \emph{CLOG-CD} ($\bigtriangleup=1$), e) \emph{CLOG-CD} ($\bigtriangleup=2$), and f) \emph{CLOG-CD} ($\bigtriangleup=4$).}
    \label{cm_CRC}
\end{figure*}

\begin{table*}[h]
\begin{center} 
\caption{Comparison between the overall classification performance of \emph{CLOG-CD} based on different oscillating steps and the baseline (ResNet-50) for each image dataset.}
\label{result_Resnet} 
\small
\tabcolsep=0.10cm
\resizebox{17.5cm}{!}{
\renewcommand{\arraystretch}{1.7}
\begin{tabular}{c || c c c c || c c c c | c  c  c c | c c c c | c c c c | c c c c } 
\hline 

Dataset      & \multicolumn {4}{c||} {Baseline} & \multicolumn {4}{c|} {\emph{CLOG-CD} (ASG)} & \multicolumn {4}{c|} {\emph{CLOG-CD} (DEG)} &\multicolumn {4}{c|} {\emph{CLOG-CD} ($\bigtriangleup=1$)} &\multicolumn {4}{c|} {\emph{CLOG-CD} ($\bigtriangleup=2$)}  & \multicolumn {4}{c} {\emph{CLOG-CD} ($\bigtriangleup=4$) }     \\ 
    & ACC & PR & RE & F1 & ACC & PR & RE & F1 & ACC & PR & RE & F1 & ACC & PR & RE & F1 & ACC & PR & RE & F1 & ACC & PR & RE & F1\\
     & ($\%$) & ($\%$) & ($\%$) & ($\%$) & ($\%$) & ($\%$) & ($\%$) & ($\%$) & ($\%$) & ($\%$) & ($\%$) & ($\%$) & ($\%$) & ($\%$) & ($\%$) & ($\%$)  & ($\%$) & ($\%$) & ($\%$) & ($\%$)  & ($\%$) & ($\%$) & ($\%$) & ($\%$) \\
\hline 
CXR  & 91.65 & 93.82  & 91.99 & 92.90 & 89.52 &90.53 &89.06 &89.79 & 93.16 & 94.84 & 94.18 & 94.51 & \textbf{96.08}  & \textbf{97.16} & \textbf{96.71} & \textbf{96.94} & 95.66 &  96.63   & 96.30    &  96.46  &  94.86  & 95.72  & 95.16 & 95.44 \\

Brain tumour & 90.89   & 89.71 & 89.83 & 89.77& 77.56  & 75.62& 71.19& 73.33&  92.20  & 91.43& 91.64 & 91.54 & \textbf{96.91}  &  \textbf{96.86}   & \textbf{96.32}   &  \textbf{96.59}  &  96.75    & 96.36    & 96.44   & 96.40 & 95.12   & 94.69  & 94.75 & 94.73 \\

knee x-ray & 63.69 & 61.36 & 62.72 & 62.03 &69.05  & 67.61& 69.19 & 68.39& 64.88  &  67.62   & 65.66 & 66.62 &  \textbf{79.76}  &  \textbf{81.60}   & \textbf{78.80}   & \textbf{80.18} &  76.19  &  77.31   & 75.65   & 76.47 & 72.02  & 74.51 & 69.05 & 71.68 \\

CRC  & 97.28 & 92.66 & 91.06 &  91.85& 97.37 & 94.46& 95.75& 96.60& 97.78 & 93.57 & 92.54 & 93.05 &  \textbf{99.17} & \textbf{99.12}  & \textbf{98.99}  & \textbf{99.06} & 98.34   & 98.34  & 98.06   & 98.20 & 98.34 & 98.11 & 98.05 & 98.08  \\
\hline
& \multicolumn{24}{c} {Without data augmentation techniques} \\
\hline
CXR  & 89.76 & 90.81 & 89.67 & 90.24 & 88.44 & 90.13& 87.05 & 88.56 & 89.75 & 90.81  & 89.67 & 90.24 & \textbf{93.58}  & \textbf{94.55}  & \textbf{94.38} & \textbf{94.47} & 88.01 & 88.60 & 88.48 & 88.54& 87.54& 87.52& 87.61& 87.57 \\

Brain tumour & 66.99 & 69.35 & 59.26 & 63.91 & 69.11& 63.95 & 62.68& 63.31 & 83.45  & 81.46 & 82.25 & 81.85 & \textbf{90.73} & \textbf{89.40} & \textbf{89.84} & \textbf{89.62} & 87.97  & 86.45  &86.11 & 86.28 & 84.88 & 82.89 & 83.20 & 83.04  \\

Knee x-ray & 35.12 & 33.84 & 35.07 & 34.44 & 58.93& 59.22 & 58.72 & 58.97 & 56.54 & 56.78  &50.26  &53.31  & \textbf{67.85}  & \textbf{69.38}  &\textbf{64.40} & \textbf{66.80} & 65.47 & 62.59 & 57.23 &59.79  & 63.69& 61.79&56.32 &58.93\\

CRC  &80.50  & 78.02& 76.37  & 77.22 & 72.47  &65.83 &65.02 & 65.42 & 85.20 & 84.09  & 82.36  & 83.22 & \textbf{88.52}  & \textbf{88.51} & \textbf{88.28} & \textbf{88.39} & 87.28 & 79.50 & 82.80 & 81.16 &83.26 &80.73 &81.88 & 81.30\\
\hline 
\end{tabular}
} 
\end{center} 
\end{table*}

\begin{table*}[h]
\begin{center} 
\caption{Comparison between the overall classification performance of \emph{CLOG-CD} based on different oscillating steps and the baseline (DenseNet-121) for each image dataset.}
\label{result_DenseNet} 
\small
\tabcolsep=0.10cm
\resizebox{17.5cm}{!}{
\renewcommand{\arraystretch}{1.7}
\begin{tabular}{c || c c c c || c c c c | c  c  c c | c c c c | c c c c | c c c c } 
\hline 

Dataset      & \multicolumn {4}{c||} {Baseline} & \multicolumn {4}{c|} {\emph{CLOG-CD} (ASG)} & \multicolumn {4}{c|} {\emph{CLOG-CD} (DEG)} &\multicolumn {4}{c|} {\emph{CLOG-CD} ($\bigtriangleup=1$)} &\multicolumn {4}{c|} {\emph{CLOG-CD} ($\bigtriangleup=2$)}  & \multicolumn {4}{c} {\emph{CLOG-CD} ($\bigtriangleup=4$) }     \\ 
    & ACC & PR & RE & F1 & ACC & PR & RE & F1 & ACC & PR & RE & F1 & ACC & PR & RE & F1 & ACC & PR & RE & F1 & ACC & PR & RE & F1\\
     & ($\%$) & ($\%$) & ($\%$) & ($\%$) & ($\%$) & ($\%$) & ($\%$) & ($\%$) & ($\%$) & ($\%$) & ($\%$) & ($\%$) & ($\%$) & ($\%$) & ($\%$) & ($\%$) & ($\%$) & ($\%$) & ($\%$) & ($\%$)  & ($\%$) & ($\%$) & ($\%$) & ($\%$)  \\
\hline 
CXR& 88.72& 90.84  &87.79 &89.29  & 86.74& 88.35 & 83.81& 86.02 & 91.69 & 88.28 & 93.76& 90.94& \textbf{94.86} & \textbf{96.10} & \textbf{95.43} & \textbf{95.76} & 93.44 & 94.94 &91.33 &93.10  & 91.27& 94.00 &90.09 &92.00
\\
Brain tumour  & 86.99 & 86.86 & 83.32 & 85.05 & 72.03 & 70.59 & 68.34  & 69.45 & 91.71 & 90.43  & 91.31 & 90.87 & \textbf{94.63} & \textbf{93.91}  & \textbf{93.99} & \textbf{93.95}  & 92.85 &92.45  & 91.65 &92.05 &91.87 & 90.91& 90.55& 90.73 \\

knee x-ray & 60.12 & 58.29 & 62.14 & 60.15 & 61.90  & 60.51 & 57.44  & 58.94  & 70.24 & 73.08  & 69.84 & 71.42 & \textbf{76.19}  & \textbf{75.59}  & \textbf{76.79} & \textbf{76.19} & 73.21 & 73.01 & 73.10 & 73.05& 72.02 & 74.81 & 71.78 & 73.23\\

CRC  & 98.20& 97.99& 97.87 & 97.93 & 96.13& 94.97 &94.23 & 94.60 &98.89 &98.83 & 98.64& 98.74& \textbf{99.45} & \textbf{99.57} & \textbf{99.40} & \textbf{99.49} & 98.34& 98.22 & 97.76 & 97.99& 97.51 & 97.66 & 96.29 & 96.97\\
\hline
& \multicolumn{24}{c}{Without data augmentation techniques}\\
\hline  
CXR & 84.38 & 85.89 & 82.66 & 84.24 & 82.16 & 84.88& 73.31 & 78.67 & 86.46 & 87.82  & 85.77 & 86.78 & \textbf{89.29}  & \textbf{90.47}  & \textbf{87.44} & \textbf{88.93} & 84.57 & 83.98 & 84.31 & 84.15 & 82.30 & 78.05 &77.90 & 77.98\\
Brain tumour  & 66.34 & 61.78 & 56.55 & 59.05 & 63.25 &56.19 & 56.91 & 56.55 & 81.63 & 80.17  & 77.19 & 78.65 & \textbf{91.87} & \textbf{90.67}  & \textbf{90.53} & \textbf{90.60} & 86.99 & 85.26 & 85.40 & 85.32& 85.37& 83.47& 83.13& 83.30 \\
knee x-ray & 39.29 & 44.09 & 33.66 & 38.18 & 44.64 & 39.12& 35.51 &  37.23& 61.31 & 61.11  & 59.60 & 60.34 &  \textbf{67.26} &  \textbf{67.15} & \textbf{64.14} & \textbf{65.61} & 60.11 & 61.19 & 58.06 & 59.59 & 55.95 & 57.05 & 53.09& 55.00\\
CRC  & 78.56 & 71.64 & 71.65 & 70.21 & 69.16 & 63.24& 60.84 & 62.02 & 87.83 & 83.90  & 84.35 & 84.13 & \textbf{92.25}  & \textbf{91.19}  & \textbf{89.60} & \textbf{90.39} & 89.76 & 87.45& 84.82 & 86.11 & 89.63 &86.95  &86.55& 86.51\\

\hline 
\end{tabular}
} 
\end{center} 
\end{table*}

\begin{table*}[h!]
\begin{center} 
\caption{Confidence Intervals at 95\% for CLOG-CD Based on Different Oscillating Steps with Baseline ResNet-50 and DenseNet-121.}
\label{CI} 
\small
\tabcolsep=0.10cm
\resizebox{16.5cm}{!}{
\renewcommand{\arraystretch}{1.7}
\begin{tabular}{c|c|c|c|c|c|c}
\hline 
method & \multicolumn{3}{c|}{ResNet-50} & \multicolumn{3}{c}{DenseNet-121} \\ 
\hline 
Dataset & \emph{CLOG-CD} ($\bigtriangleup=1$) & \emph{CLOG-CD} ($\bigtriangleup=2$) & \emph{CLOG-CD} ($\bigtriangleup=4$) & \emph{CLOG-CD} ($\bigtriangleup=1$) & \emph{CLOG-CD} ($\bigtriangleup=2$) & \emph{CLOG-CD} ($\bigtriangleup=4$) \\ 
 
\hline
CXR & (94.42\% and 95.31\%) &  (94.08\% and 94.85\%)   & (93.92\% and 94.41\%) & (88.08 and 92.56) & (86.79 and 90.83) &(80.29 and 87.06)\\ 
\hline
Brain tumour & (94.09\% and 95.69\%) &(91.00\% and 94.16\%) &  (90.19\% and 93.08\%) & (84.07 and 91.35)& (76.47 and 90.30)&(73.04 and 88.72)\\  
\hline
Knee x-ray & (74.42\% and 77.90\%) & (69.66\% and 73.61\%) &  (64.78\% and 67.99\%)  & (69.08 and 75.21)&(68.71 and 72.23) & (67.44 and 70.79)\\  
\hline
CRC  & (84.65\% and 95.34\%) & (83.74\% and 93.22\%) &  (79.69\% and 91.41\%) & (96.77\% and 99.77\%) & (94.17 and 98.01)& (87.08 and 96.49)\\ 

\hline 
\end{tabular}
} 
\end{center} 
\end{table*}

\subsection{Comparison with state-of-the-art methods}


Finally, we compared our proposed method with two different CL-based methods: DCLU \cite{li2023dynamic} and curriculum learning with prior uncertainty \cite{jimenez2022curriculum}, both evaluated without using the AUG technique. As shown in Table \ref{comp}, the overall classification accuracy of DCLU was significantly lower than that of \emph{CLOG-CD} on the CXR and digital knee x-ray datasets, while it was slightly lower on the brain tumour and CRC datasets. Furthermore, compared to the curriculum learning with prior uncertainty method \cite{jimenez2022curriculum}, our proposed method outperformed the results across all datasets.

To further highlight the effectiveness of our \emph{CLOG-CD} method, we compare its performance with several previous works using the same datasets but with different experimental settings. In the case of CXR classification, \emph{CLOG-CD} achieved an accuracy of 96.08\%, outperforming CNN-DenseNet201 (95.11\%) \cite{rahman2021exploring}, CNN-LSTM (94.50\%) \cite{naseer2023deep}, and \emph{CoroDet} (94.20\%) \cite{hussain2021corodet}. Similarly, for brain tumour classification, \emph{CLOG-CD} achieved a high accuracy of 96.91\%, surpassing techniques such as CNN-MobileNetV2 (92.00\%) \cite{tazin2021retracted}, Genetic Algorithm (94.34\%) \cite{noreen2021brain}, the 7-layered CNN (84.19\%) \cite{abiwinanda2019brain}, and \emph{XDecompo} (94.30\%) \cite{abbas2022xdecompo}. In the classification of digital knee x-ray images, \emph{CLOG-CD} achieved an accuracy of 79.76\%, outperforming CNN-ResNet-50 (64.58\%) \cite{liu2022novel}, CNN-VGG-19 (69.70\%) \cite{chen2019fully}, and CNN-LSTM (75.28\%) \cite{wahyuningrum2019new}. Finally, in the CRC dataset, \emph{CLOG-CD} achieved the highest accuracy of 99.17\%, outperforming ICAL (94.07\%) \cite{hu2023learning}, the Multi-class texture with CL (94.30\%) \cite{kather2019predicting}, and the Multitask ResNet-50 model (95.0\%) \cite{peng2019multi}.\\

\begin{table}[h!]
\begin{center} 
\caption {Comparison of the \emph{CLOG-CD} model with curriculum-based methods. }
\label{comp}
\small
\tabcolsep=0.10cm
\resizebox{8.5cm}{!}{
\renewcommand{\arraystretch}{1.0}
\begin{tabular}{c|c|c|c|c}
\hline
&\multicolumn{4}{c}{ACC (\%)}\\
Method  & CXR  &  brain tumour  & knee x-ray  &  CRC  \\
\hline
DCLU \cite{li2023dynamic}   & 88.44 & 90.20 & 49.40 &  91.84  \\
CL with prior uncertainty \cite{jimenez2022curriculum}   & 61.11 & 51.38 & 40.00 & 65.28 \\
CLOG-CD (ResNet-50)& \textbf{93.58} & 90.73 & \textbf{67.85} & 88.52\\
CLOG-CD (DenseNet-121)& 89.29 & \textbf{91.87} & 67.26 & \textbf{92.25}\\
\hline
\end{tabular}
}
\end{center} 
\end{table}

\section{Discussion and conclusion}
\label{discussion}

The curriculum learning strategy is a powerful machine learning process that aims to enhance model performance by achieving faster convergence. However, in the medical image domain, datasets often suffer from irregularities in class distribution, which can significantly impact model performance. Fortunately, the class decomposition approach has proven to be a robust solution to this problem. In this paper, we introduce a new CNN training approach called \emph{CLOG-CD}. To the best of our knowledge, this is the first attempt to employ class decomposition within the curriculum learning strategy for medical image classification.
The proposed method leverages the learned weights of the CNN from different levels of granularity of decomposition to serve as the CL scheduler, with the levels of decomposition granularity ranked in descending-ascending order. \emph{CLOG-CD} model has demonstrated its ability to improve model performance on target tasks and accelerate the training process, especially with the datasets suffering from irregularities within the classes, such as CXR, brain tumour, digital knee x-ray, and CRC datasets. Firstly, we investigated the effectiveness of \emph{CLOG-CD} with a single speed step in one direction, where the levels of decomposition granularity were ranked in descending order \emph{CLOG-CD} (DEG). The results demonstrated that the \emph{CLOG-CD} (DEG) model outperforms traditional curriculum learning, \emph{CLOG-CD} (ASG), as well as traditional transfer learning with baseline models, ResNet-50 and DenseNet-121. This improvement arises from simplifying complex tasks from the beginning through class decomposition, enabling the model to focus on capturing essential features and efficiently learning complex patterns.
By addressing the complexity tasks first, the model becomes more effective to handle the dataset's irregularities and complexities, ultimately leading to improve the classification performance. Secondly, we evaluated the \emph{CLOG-CD} model in both directions and with different curriculum oscillation processes. The experimental results showed that the \emph{CLOG-CD} ($\bigtriangleup=1$) model achieved the highest classification performance on the test sets of all datasets after 50 epochs, compared to other training strategies. Using ResNet-50 pre-trained network, our proposed method achieved an ACC of 96.08\% for CRC, 96.91\% for brain tumour, 79.76\% for digital knee x-ray, and 99.17\% for CRC datasets. Moreover, we evaluated the performance of our method with another baseline, DenseNet-121. The results also demonstrated a significant improvement in classification performance of \emph{CLOG-CD} ($\bigtriangleup=1$) model compared to other training strategies, achieving 94.86\%, 94.63\%, 76.19\%, and 99.45\% for CXR, brain tumour, digital knee x-ray, and CRC datasets, respectively.
In addition, Table \ref{comp} demonstrates that our method has achieved significantly better performance compared to other curriculum-based methods.



\bibliographystyle{IEEEtran} 
\bibliography{references.bib}

\end{document}